\documentclass[preprint,12pt]{elsarticle}

\usepackage{amssymb}
\usepackage{amsmath}
\usepackage{enumitem}

\journal{Nuclear Physics B}

\begin{document}

\begin{frontmatter}



\title{Blockchain-IoT Convergence for Spatiotemporal Data Systems: Challenges, Solutions, and Research Gaps}

%

 \author{Natalia Chaudhry}
 \ead{natalia@pucit.edu.pk}

 \author{Muhammad Murtaza Yousaf}
 \ead{murtaza@pucit.edu.pk}
 \affiliation{organization={Department of Software Engineering},Department and Organization
             addressline={University of the Punjab},
             city={Lahore},
             postcode={}, 
             state={Punjab},
             country={Pakistan}}

\begin{abstract}
Blockchain in conjunction with the Internet of Things (IoT) is changing the management of spatiotemporal data in a way that has never been possible before now. In the process of gathering geospatial and time-stamped data from a large network of IoT devices, data authenticity, accuracy, and timely processing are major issues. Spatiotemporal indexing in a blockchain environment is necessary for ensure quick data retrieval; supporting real time data processing requirement. Blockchain with its decentralized consensus mechanism is a possible solution for preventing data fabrication, unauthorized access, and single point of failure in IoT based spatiotemporal applications. This paper aims to describe how Blockchain and IoT indexing can be combined to enhance the spatiotemporal data management (specifically indexing and querying of data). Literature solutions will be reviewed and compared to highlight challenges and insights for future blockchain-integrated IoT in resilient spatiotemporal data ecosystems.
\end{abstract}


\end{frontmatter}




\section{Introduction}
\begin{figure}[ht]
    \centering
\includegraphics[width=8cm,height=9cm,keepaspectratio]{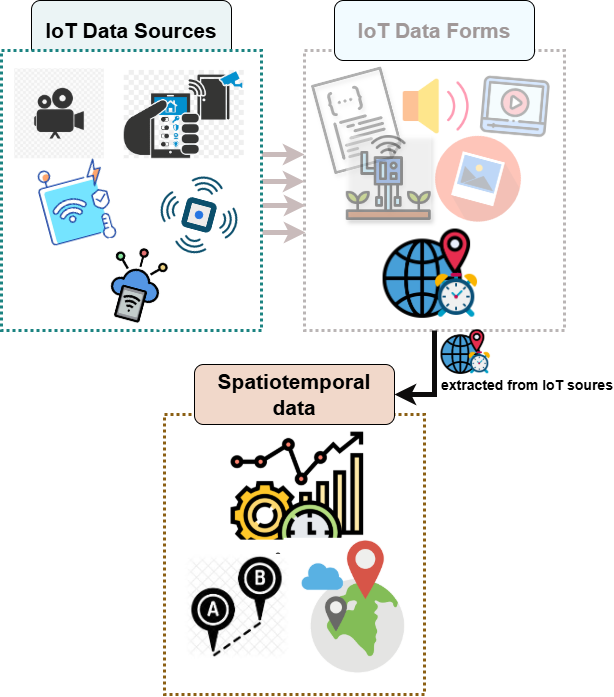}
    \caption{IoT data sources and data formats}
    \label{iot}
\end{figure}

Due to the rapid development of Internet of Things (IoT), there is an increasing amount of data that is spatial and temporal in nature. IoT sensors generate variety of data including textual data, images, multimedia, sensory data, spatiotemporal data. The focus of this paper is primarily on spatiotemporal data as illustrated in Figure \ref{iot}. The secondary focus is on blockchain enabled solutions that are based on general IoT data indexing or index targeting either spatial or temporal aspects. 
Smart cities, intelligent transportation systems, environmental monitoring, and supply chain logistics are some of the areas that rely on IoT devices that produce huge amounts of geospatial data in real time, which needs to be stored, secured and scaled up efficiently. However, guaranteeing the accuracy, protection and the real time delivery of such data in a highly decentralized environment is still a major issue. Blockchain technology could be a solution to this problem as it brings decentralization, immutability and cryptographic security to IoT systems. By using blockchain technology, the spatiotemporal data produced by IoT is securely stored and shared, and its integrity can be proved without the help of a third party. Smart contracts can also be used to facilitate the sharing of data between different participants based on trustless rules. However, for the integration of blockchain with IoT to become mainstream, the issues of scalability, energy consumption, and real-time data processing must be solved. Although the application of blockchain-enabled IoT has been identified in different sectors, the management of spatiotemporal data requires particular mention. The integration of indexing in such environments is mandatory to address the challenges of efficiency, scalability, and real time query processing. IoT sensors generate enormous amount of data that needs indexing solutions for faster querying and retrieving of time and location based data. As these sensor devices grow in their count, the volume of data also increases along. It is important to reduce search complexity using indexing schemes and ensure scalability by optimizing data retrieval and storage. This is useful in many IoT applications that have real time processing demands such as smart traffic management and tracking and autonomous vehicle tracking.

Blockchain technology has emerged as a revolutionary breakthrough in a wide range of areas with spatiotemporal applications. There are a number of factors that contribute to the significance of blockchain in spatiotemporal applications. Some of them are outlined as under.
Decentralization: Blockchain architecture eliminates the need for central authority to manage data storage and processing. In spatiotemporal applications, this feature supports distributed data storage ensuring resilience and fault tolerance. This is specifically useful for applications requiring real time access to data.
Immutability: Blockchain involves an immutable ledger in which transactions are not altered. This ensures integrity of spatiotemporal data specifically in applications where data provenance is critical.

\textbf{Security:} Blockchain uses cryptographic techniques to secure transactions from unauthorized access. This is critical in spatiotemporal applications that deal with sensitive information such as data related to asset tracking, location, and environmental monitoring. Cyberattacks and data breaches can be avoided.

\textbf{Transparency and Trust:} Through the provision of a shared and transparent representation of data and transactions, participants besides benefiting from transparency can place their trust on the technology using Blockchain. In spatiotemporal applications fostering collaboration and cooperation like government agencies business and citizens, this transparency plays a vital role. Blockchain can, for instance, ensure transparent and accountable governance in the smart city programs by giving its audience the ability to access as well as check any data in relation to public entities.

\textbf{Smart contracts:} Most blockchain platforms support smart contracts. Smart contracts are self-executing contracts that have predefined conditions or rules. Spatiotemporal data aspects of contracts include data sharing agreements, access permissions, and data monetization laws. Such automation process is used in order to increase efficiency of spatiotemporal applications and also quality assurance through reduction of human errors while at the same time ensuring that contractual obligations are met.

\begin{figure}
    \centering
    \includegraphics[width=0.8\linewidth]{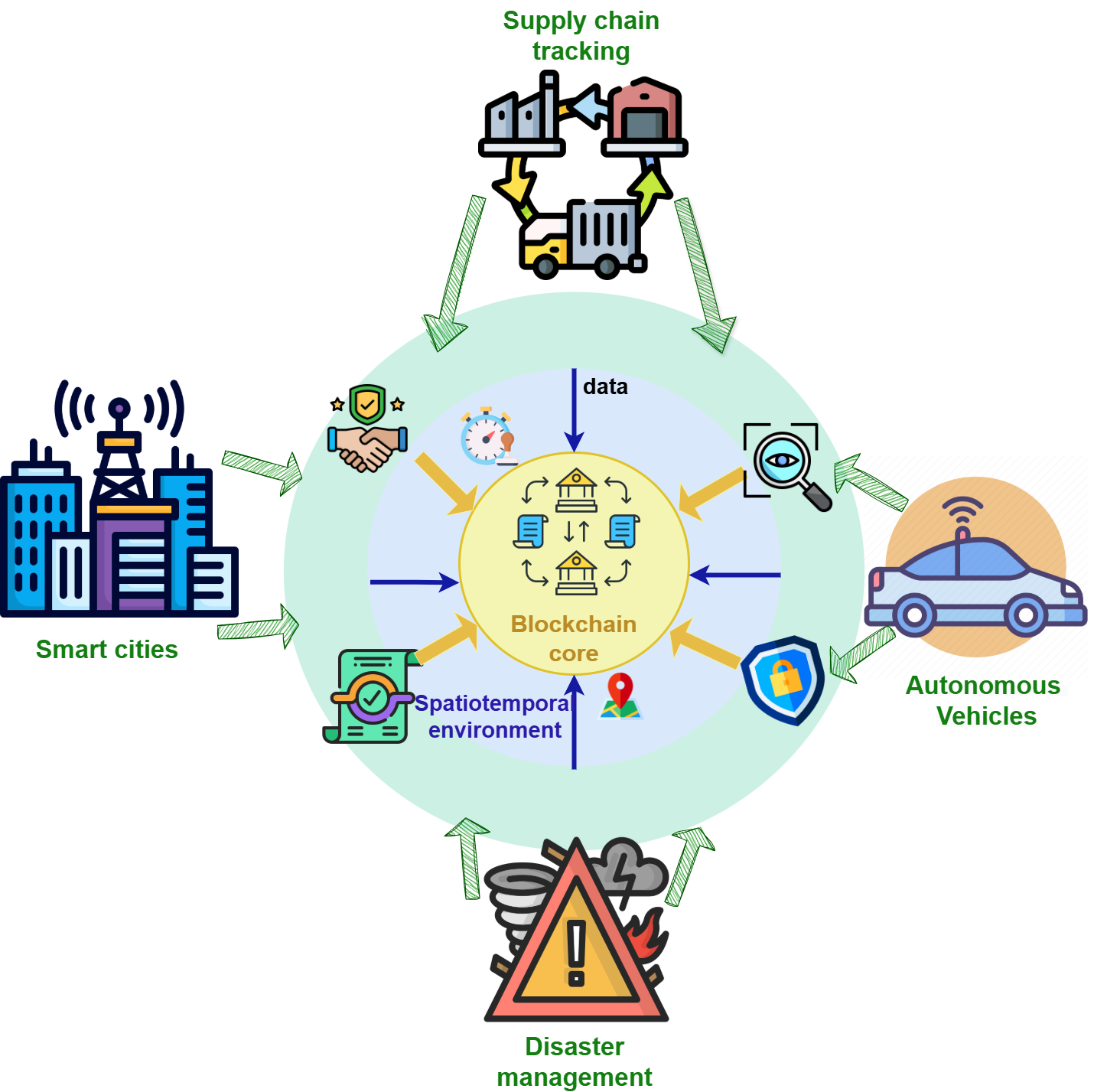}
    \caption{Blockchain and Spatiotemporal Data Fusion: Enabling Security, Transparency, Trust, and Smart Contracts}
    \label{f1}
\end{figure}

Blockchain technology has immense potential for spatiotemporal applications. This is because it provides a secure infrastructure that is both transparent and decentralized while at the same time being able to manage geospatial as well as temporal data. What this means is that by taking advantage of some of the most important aspects of blockchain which include decentralization, immutability, security, transparency and smart contracts, there are possibilities available to be exploited by companies and societies. These could be in areas like supply chain management, smart grids, healthcare among other things. Figure \ref{f1} illustrates the fusion of blockchain with spatiotemporal data management and applications.
The integration of blockchain technology and IoT applications introduce both challenges and opportunities outlining some grey areas that need to be addressed. Blockchain can help in improving the response time of queries but designing an index structure for complex queries in IoT environment involving various time, data and location dimensions is challenging and needs innovative solutions. Ensuring privacy of data my making indexed data accessible only to authorized users while maintaining the immutability of blockchain is also a grey area that is worth considering. Optimization of performance when dealing with huge volume of data generated by IoT devices without compromising scalability is critical. Another grey area lies in designing standardized index structures that can withstand the heterogeneity of data formats generated by IoT sensors. Addressing these challenges and grey areas need multidisciplinary approach that take into consideration ethical, technical, and performance implications. IoT systems may have to deal with range queries based on multiple dimensions of data i.e. time, location, and data type. Blockchain only supports queries on transactions using timestamps that makes the query efficiency low. Many research studies exist that tried to improve query efficiency by making amendments to the structure of blockchain. 

To the best of our knowledge, none of the studies in the literature have reviewed the fusion of blockchain and spatiotemporal data indexing and management in detail. The existing studies are mostly based on only blockchain concept and comparisons have been made with respect to blockchain specific features. However, our paper discusses, reviews, and compares the efforts made in the literature regarding fusion of both the domains (blockchain and spatiotemporal data management and indexing). We have reviewed in detail the research works that have either designed a query or indexing method for spatiotemporal blockchain enabled environment or presented a framework. We also discussed the suitability of solutions in real life applications. Challenges along with meaningful insights have also been presented in this paper which we believe is missing in the existing surveys or comparative studies.

We first identified and discussed the features and aspects that are significant to discuss with respect to spatiotemporal data management. It is worth to mention that data management is a broader term with many aspects beyond indexing and query processing such as Data partitioning, data security, ETL and other integration services, data analytics, backup/recovery, data collaboration, cloud data management. However, during survey we focused only on studies that have either proposed an indexing structure, data management framework, or query processing algorithm. Thus, primary aspects covered under data management are data indexing and data query processing. We assessed how blockchain can improve location and spatial analysis, as well as real-time decisions in IoT applications. We explored the existing works on these terms in detail and comparative analysis is also performed. The main contributions are:
\begin{itemize}
    \item We reviewed the research works in the literature focused on the integration of blockchain and IoT data management; with special focus on spatiotemporal data indexing and querying. We reviewed the works that have either presented a query or indexing method or a framework for blockchain enabled spatiotemporal environment. 
    \item We identified and presented some important aspects and features that need attention while reviewing any spatiotemporal and blockchain enabled solution. We grouped the features into relevant categories (feature vectors) for the sake of clarity and better understanding.
    \item Challenges in such environment are also highlighted and discussed in detail. Existing comparative studies lack the discussion on challenges especially for spatiotemporal enabled blockchain environment.
    \item We also assessed the suitability of existing solutions (present in the literature) in real life applications. Existing surveys and studies have not discussed real life applications during analysis and comparisons. 
\end{itemize}

As per our knowledge, none of the existing research works in the literature have performed such extensive comparative analysis specifically for spatiotemporal data. Rest of the paper is organized as follows. Section 2 discusses the work done in the literature on integration of two technological advancements: IoT (specifically spatiotemporal data) and blockchain. Section 4 presents detailed comparative analysis according to the features and feature vector discussed in Section 3. Section 5 presents meaningful insights for optimizing blockchains and spatiotemporal data management. Section 6 concludes the study.
\section{Methodology for Achieving Research Objectives}
A structured approach is followed for literature review to ensure transparency and reproducibility of research. The studies relating to blockchain and spatiotemporal indexing in an IoT enabled environment are identified and evaluated. Following electronic databases are considered during literature survey.
\begin{itemize}
    \item Scopus
\item IEEE Xplore
\item ACM Digital Library
\item Google Scholar
\end{itemize}
This allowed coverage of high impact journals on interdisciplinary research involving blockchain, spatiotemporal data management and indexing, and IoT. The search strings used in the literature search, as defined below, forms the Boolean combinations of predefined keywords aligning with the research theme and objectives. 

(Blockchain or distributed ledger) AND data management OR data indexing OR data structure OR query processing AND (spatiotemporal OR IoT)

The search is restricted to conference and journal articles, written in English, and published in the period of 2018-2025. The including/excluding criteria for research studies is discussed as under. 

Include studies that have:
\begin{itemize}
    \item Either proposed or studied blockchain based data management systems.
\item Studied or proposed index structures for blockchains
\item Investigated query processing with regard to spatiotemporal environment and blockchains.
\item Investigated IoT enabled blockchain based data management systems or approaches.
\end{itemize}
Studies were excluded if they:
\begin{itemize}
    \item Do not relate directly to blockchain based data management.
\item Seem duplicated across databases
\item Do not focus on data management 
\item Do not have full text available
\end{itemize}
The article screening procedure is inspired from PRISMA \cite{selccuk2019guide} procedure with stages as defined below.
\begin{itemize}
    \item \textbf{Identification:} Initial articles are retrieved using the defined search criteria.
\item \textbf{Deduplication}: Duplicate articles are removed/dropped.
\item \textbf{Screening}: Articles’ title and abstract are studied to screen them further with respect to relevance to blockchain based spatiotemporal data management and indexing.
Full text review: Remaining articles are assessed in totality to find out eligibility.
\item \textbf{Final inclusion}: Studies and articles that meet the defined inclusion criteria are finally incorporated into literature analysis. 
\end{itemize}

We found only 20 relevant research articles as the research on fusion of spatiotemporal data management and blockchain is still in its initial progressing phase. For detailed evaluation, we defined grouping of the retrieved research studies. We divide the literature into two groups: 1) surveys, and 2) proposal of either an index or framework or query processing solution. Summarized view is presented in Table \ref{tab7}. 
\begin{table*}[ht]
    \centering  
    \begin{tabular}{|p{2.5cm}|p{6cm}|p{2cm}|}
    \hline
     \textbf{Category}  &   \textbf{Research works included in category}  &  \textbf{Count} \\
       \hline
Proposals & \cite{zheng2024data}, \cite{hojati2024dstree}, \cite{xing2021blockchain}, \cite{jiang2019cross}, \cite{chen2020data}, \cite{nasrulin2018robust}, \cite{qu2019spatio}, \cite{wen2021blockchain}, \cite{li2021blockchain}, \cite{trimborn2018crix}, \cite{liu2019energy}, \cite{feng2022efficient}, \cite{zhang2019gem}, \cite{lee2023hierarchical} & 14
\\
      \hline   
Surveys & \cite{lu2018blockchain}, \cite{liu2020blockchain}, \cite{ferrag2018blockchain}, \cite{elmamy2020survey}, \cite{chitti2019data}, \cite{xiong2020best} &  6 \\
      \hline 
    \end{tabular}
    \caption{Summarized view of research articles}
    \label{tab7}
\end{table*}
\section{Related work}
\begin{figure}[ht]
    \centering
    \includegraphics[width=0.7\linewidth]{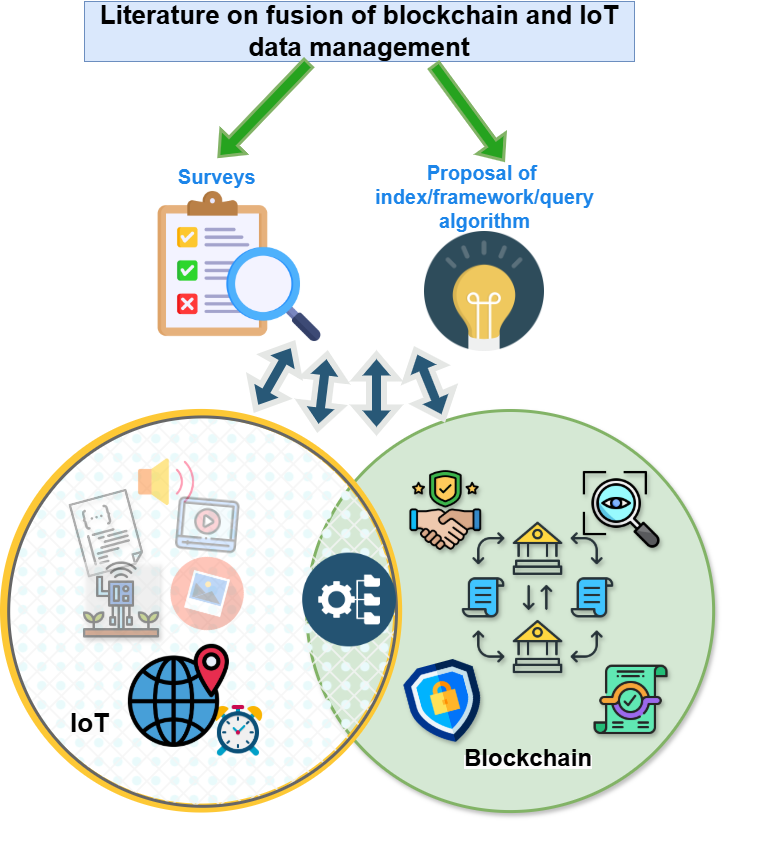}
    \caption{Literature survey taxonomy}
    \label{lit}
\end{figure}

The related work in terms of defined grouping is discussed as under. The goal is to incorporate such studies in comparative analysis that have either presented a new query processing method, or a new data structure or indexing solution, or a framework for a blockchain enabled environment, focusing on IoT (in general) and spatiotemporal data management (in specific). Figure \ref{lit} presents a scope of this study indicating special focus on spatiotemporal data indexing and querying aspects along with blockchain integration.
\subsection{Surveys and comparative studies}
The paper \cite{lu2018blockchain} reviewed the current status of challenges in using blockchain and also provided a number of future research directions. In depth assessment of evolution of blockchain, applications of blockchain and its features is performed. Implementation challenges re also stated in terms of data privacy and management. Various consensus mechanisms such as Proof of Work, Practical Byzantine Fault Tolerance, and Proof of Stake are discussed; explaining how integrity and security is achieved using them.

The paper \cite{liu2020blockchain} studies the integration of blockchain and data science; highlighting the relationship of blockchain with big data and security challenges. The integration has resulted into a number of benefits such as improved data access, security, and sharing. Open issues and challenges are also identified.

Survey of blockchain protocols suitable for IoT network are reviewed in the paper \cite{ferrag2018blockchain}. Threat models are classified into five categories for better analysis. These are manipulation, reputation, identity, service oriented and cryptanalytic attacks. Research challenges are also discussed along in detail.

The paper \cite{elmamy2020survey} reviewed contributions made in the literature on cybersecurity and blockchain. Classification of significant cyber attacks based on four categories within the context of industry 4.0 is proposed. These categories are local/remote attacks, denial of service attacks, power of root attacks and scanning attacks. Existing literature on blockchain in Industry 4.0 are compared focusing on confidentiality, integrity, and availability. Open research problems relating to performance and security of blockchain in Industry 4.0 are outlined.

A comparative analysis of blockchain databases and relational databases has been performed in \cite{chitti2019data}, highlighting the strengths and weaknesses of each technology. The work outlined the features of each technology and provides guidance to developers in selecting suitable database for their application. The potential for integration of technologies is also discussed. BigchainDB is specifically analyzed in this respect that combines both the technologies.

The paper \cite{xiong2020best} enforces the use of blockchain in decentralized data management to assume the benefits of blockchain technology such as auditability, transparency, and traceability. A resource allocation method is introduced that optimizes data transmission among IoT devices in the blockchain environment. A performance effective stochastic optimization framework is presented for channel selection during transaction transmission.

\subsection{Contributions relating to blockchain, enabled IoT data indexing}
The paper \cite{zheng2024data} proposed a dual-layered architecture named S-DAG for managing data in IoT systems. It combines directed acyclic graph and blockchain to avoid query inefficiency and overheating issues. The adaptive load balancing algorithm balances the performance of nodes. Three-dimensional queries are entertained with a 3D-Merkle index structure that combines KD tree and Merkle tree that improved the efficiency of transactions. However, the solution is complex to implement and has resource requirements. Edge nodes can handle spatiotemporal data to enable advanced analytics on spatiotemporal information that is useful in monitoring and predictive decision making. Edge node is a device that connects IoT devices and cloud server. The edge nodes lie near to where data is generated and perform tasks of data aggregation, filtering, and processing before sending it to the cloud for further processing. The paper also evaluated existing solutions of sharding. ABLA algorithm is proposed that takes into consideration both the transaction processing time and hardware oriented capabilities (in terms of edge nodes) to deal with nodes that have low load thus preventing overheating of shards. It performs resharding and nodes adjustment to balance the performance load. 

The paper \cite{hojati2024dstree} presented a framework based on blockchain technology for managing two dimensional data supporting vector based data in distributed networks. An indexing structure for spatio-temporal data is proposed to deal with range queries. Two primary parts of index are interval tree and quadtree. Interval tree deals with point queries handling overlapping time intervals. Quadtree deals with spatial aspects of queries. The focus is primarily on range queries, topological queries, and temporal queries. The proposed structure deals only with append only data and is not optimized for deletion operations. This restricts it to scenarios in which frequent deletions are involved.  
Sharding is not discussed explicitly however, data management in distributed environment is discussed. This implies scalability and efficient query processing without depending on centralized systems. The distribution across peer to peer networks allows to scale the system as increase in data volume introduces bottleneck issues. The paper does not explicitly discussed energy consumption however, several aspects influence energy efficient consumption. Local query processing at replicated versions of index minimizes data transmission across remote servers. Further, each P2P node manages its own data avoiding centralized bottleneck issues leading to minimization of energy costs across the network. The paper discusses smart contract usage for data access control on the blockchain. The consensus mechanisms are not explicitly discussed in this paper.

The paper \cite{xing2021blockchain} presented a novel structure named SCATC that simplifies the query processing by dividing the transaction chain into subchains; thereby reducing the storage capability and enhances scalability particularly for transactions with extensive transaction histories. It merges multiple transactions with same account into the same block and also shortens the path of query related to account subchain data. This reduces the data accesses to irrelevant block. Hash pointers are used to connect subchains. Sharding is not explicitly discussed. Energy consumption is not discussed however, enhanced query performance may indirectly lead to lower power consumption. The solution presented in this paper does not ensures data privacy nor it uses any consensus mechanism that is quite important for maintaining agreement in distributed blockchain environment. 

A scalable framework is proposed in \cite{jiang2019cross} to integrate multiple blockchain platforms for efficient management of IoT data generated by sensors. The framework supports parallel verification of transactions. The paper does not specifically address spatiotemporal data. Energy consumption is not explicitly discussed but the use of light weight clients may indirectly lead to lower consumption of energy. The paper discussed the use of BigChainDB, IPFS and DAG structure for storage and efficient transaction processing. Data access control protocol and cryptographic measures ensures security of transactions. Byzantine Fault Tolerance (BFT) consensus algorithm is used for consensus. Due to parallel processing in Tangle, throughput is reported to be increased leading to reduced latency. Queries through smart contracts are supported by the proposed framework. The use of replication strategies employed is not discussed in detail. However, consortium blockchain implies that some replication mechanism exist that replicates data across the nodes to ensure fault tolerance. The BFT mechanism ensures strong consistency. 

The paper \cite{chen2020data} introduces a data management method that ensures decentralized management, trust based data exchange, tamper proof data, and data privacy. A non-temper index utilized by BigchainDB and blockchain structure with hash pointers is discussed that facilitates in quick retrieval. Sharding and energy consumption is not explicitly discussed. BigchainDB is mentioned to have a alliance voting based consensus mechanism that aids in achieving scalability. Some other consensus mechanisms such as Byzantine fault tolerance and proof of work are also reviewed. Security's significance is also discussed to avoid vulnerabilities in smart contracts. Blockchain query mechanisms are discussed in generic way with no specific query types mentioned or discussed. The focus on consistency and consensus reflects awareness about concepts of CAP and DCS. However, no specific discussed is made regarding these concepts. Consistency and consensus are both discussed and emphasized with regard to their importance. Handling of spatiotemporal data is not discussed explicitly. 

The paper \cite{nasrulin2018robust} proposed a proof-of-location protocol for verification of spatiotemporal data dealing with location attributes. It also presented a sharding strategy that improves load balancing and provides scalability. Security protocols are also integrated into the proposed framework to protect against unauthorized access to data. A threat-to-validity model is presented in this regard. Robust replication strategy which is crucial for fault tolerance is introduced. Consensus algorithm is proposed that improves synchronization across nodes of the network; tailored to check for spatio-temporal data validity. Replication is not discussed explicitly nor any new replication model is proposed; however, permissioned blockchain employed in this paper inherently provides replication. Tradeoffs among consistency, availability and partition tolerance is discussed and distributed consensus concept is introduced to balance these aspects. Eventual consistency models are emphasized to address consistency concerns. Since permissioned blockchains ensure strong consistency using mechanisms like practical Byzantine Fault Tolerance, the solution proposed in this paper inherits these consistency properties.

Integration of blockchain and data management practices has been assessed in the paper \cite{paik2019analysis}. Using the lens of data management, blockchain technology has been examined that highlights the insights that are crucial for software designing. Typical layers of software system are conceptualized and it is examined how blockchain fits into it. Data administrative aspects are also discussed to assess role of blockchain in distributed data store management and future research directions are presented. 
 
Three indexing structures for immutable data; Merkle Bucket Tree (MBT), Pattern-Oriented-Split Tree (POS-Tree), and Merkle Patricia Trie (MPT) are comprehensively evaluated in the paper \cite{yue2020analysis}. While evaluating the indices, both theoretical and empirical performance characteristics are taken into focus including the worst case performance guarantee of each index. POS-Tree has been reported as the most favorable candidate in indexing immutable data.  

Critical issues and challenges in data management are identified in \cite{wen2019application} and it is also assessed how blockchain addresses them. Traceability, security, decentralization, automatic verification, and data consistency are discussed as primary advantages of blockchain. The paper also presented a data management model based on blockchain to make the data management procedure more effective by incorporating authentication, encryption algorithms, consensus mechanisms, verification and sharing of data.

The paper \cite{qu2019spatio} presented a novel query processing solution to entertain spatio-temporal data. Algorithm named as "Traversal for Generalized Spatio-temporal blockchain index" is presented that can deal with queries running over block-DAG structure. Detailed evaluation of proposed solution is conducted that confirms that practical applicability of the solution to real life scenarios. The proposed solution is scalable and can handle large datasets and variety of queries including ball-point queries, bounded k-NN queries, k-NN queries and range queries. The algorithm is expected to provide consistent results by cryptographically signed history in block-DAGs. The algorithm is shown to outperform the traditional blockchain systems however, exact numeric values have not been reported. Same is the case with latency. The results indicate that latency is directly proportional to the number of nearest neighbors (for k-NN queries). For ball point and bounded k-NN queries, increasing the bounding radius does not fluctuates the overall query time that indicates that the solution can handle large search areas without compromising the performance and increasing latency.

A close contact tracing solution is proposed in \cite{wen2021blockchain} that uses blockchain to utilize spatio-temporal information prioritizing user privacy. Close contact tracing queries are supported by the proposed system dealing with spatiotemporal data. The system is designed to be scalable with throughput of 100 transactions per second. The solution employed used Geohash and Unix timestamps for geographic and temporal indexing. Using smart contracts for access control and encryption mechanisms amplifies the security of the system.

The paper \cite{li2021blockchain} proposed a blockchain enabled data collection framework using UAV assisted wireless sensor networks. Sparsity-optimized and compressed sensing approach (SCSDA) utilized in this work is used to reduce data redundancy; ensuring energy efficiency. The proposed solution uses sparsity optimized aggregation technique that enhances the scalability. Data reconstruction-directed consensus mechanism is proposed that accurately reflects the data collected from sensory devices; thereby enhancing the reliability of the system.

The paper \cite{trimborn2018crix} presented an index that captures the dynamics of cryptocurrency market and can adapt to market changes. The paper does not indicate a specific data structure or model utilized.

Traditional methods for query processing in WSN rely on a fixed structure or network topology that results into high energy consumption and inefficiencies. The paper \cite{liu2019energy} proposed a dynamic query processing algorithm that is not dependent on a pre-defined network structure. This design minimizes energy consumption to a great extent by allowing only specific nodes in the network to participate in aggregation query processing. Privacy protection method of sharing data is also proposed that upgrades the security of data aggregation. The method deals with dividing the data into multiple chunks (shares) thereby ensuring data sharding. Existing aggregation query protection algorithms are also reviewed. 

A new retrieval system for decentralized environment has been proposed in the paper \cite{feng2022efficient}. A secondary index (BB+ tree) based on B+ tree and Bloom filters has also been proposed to serve retrieval operations. Time attributes of blockchain transactions is managed with a newly introduced index named as I2B+ tree to further enhance the performance. The proposed indices gave considerable performance improvement, with significant reduction in retrieval time of single and multi feature retrieval queries. However, these indices structure consumes extra space which negatively effect scalability if the number of accounts grow significantly. Although Bloom filter helps in making the system space efficient however, they might introduce false positives that leads to unwanted lookups in B+ tree. Further, security implications have also not been discussed.

The paper \cite{zhang2019gem} introduced a data structure called GEM2-tree to deal with authenticated range queries for a blockchain. The structure is made to work under a gas cost model that aims at minimizing transactions costs. The proposed method resulted in reduction in gas costs by a factor of 4, in comparison with traditional methods. To enhance scalability, the paper introduced a two level index that deals with large datasets. The index relies on hash values to ensure integrity and consistency of data.

A structure (hierarchical quadrant spatial LSM tree (HQ-sLSM tree)) for indexing geospatial data from blockchain system is proposed in \cite{lee2023hierarchical}. Te data is first linearized using geohashes before inserting into index. To process range and nearest neighbor queries, spatial filtering is applied that ensures low disk I/O operations. The focus of this paper is on enhancing insert performance to make the solution feasible for blockchain environment dealing with write-intensive operations. The index contains components for both memory and disk. Memory component deals with recently inserted data that allows for quick access in write intensive environment. Once the capacity of exceeds a certain threshold, four quadrants space is formed and data is flushed to disk. The index does not directly deal with spatiotemporal data. Its primary focus is on spatial indexing. HQ-sLSM tree supports range and nearest neighbor queries and optimizes them using MBR usage, query filter generation, spatial filters, effective result management, and selective component access. 

To summarize, blockchain when used in IoT applications offers several advantages such as integrity of data, enhanced data security, and decentralized trust. Along with these benefits, some limitations are also worth mentioning. When a large number of IoT devices generate tremendous amounts of transactions, the blockchain networks face a scalability crisis due to inherent consensus and replication mechanisms. The consensus procedure instigate latency which results in delayed response in IoT applications. The computational power, storage, and bandwidth of IoT sensors is usually limited. Adopting blockchain technology and maintaining distributed ledger for IoT applications can be challenging as the devices are resource constrained. Furthermore, for IoT deployments, the network and computational expenses of blockchain may become significant and infeasible to bear. In case of public blockchain, data confidentiality and privacy concerns may arise for storing IoT data. Heterogeneity is also a major concern in utilizing blockchains. IoT applications usually deal with diversified communication protocols and varying data formats. Integration of these devices and blockchain networks requires careful standardization of solutions to ensure interoperability and seamless exchange of data. Mining operations in blockchain also consume a considerable amount of energy which makes it difficult to align with energy-efficient requirements of IoT devices. Addressing these challenges needs careful inspection of requirements of IoT deployment, consensus and data management strategies, blockchain platform’s selection to realize the benefits of blockchain technology while addressing these limitations in IoT environments.

\section{Features for comparative analysis}
The literature on fusion of blockchain and IoT spatiotemporal data management is explored in detail. To the best of our knowledge, none of the existing surveys have reviewed the literature works and compared them in detail with respect to multiple aspects; including structural, performance oriented, and other significant aspects. The existing works have either compared the works briefly with respect to few basic aspects of blockchain like immutability, security, and consensus. However, since we aim to study the fusion of two significant sub domains (IoT and spatiotemporal data management) and blockchain; it is very important to come up with detailed analysis covering aspects of both the sub domains.  

During literature analysis we extracted some useful features that are relevant to consider while evaluating and comparing the research works. These are outlined and briefly described in section defined under.
\begin{itemize}
    \item \textbf{Sharding}: Sharding  a technique that leads to enhancing scalability of blockchain by breaks down the data into small manageable pieces known as shards. Each shard can then process the data in parallel; thereby increasing efficiency and throughput of the system. It assists in eliminating network congestion by distributing transactions into manageable shards. This reduces the load on a single node leading to faster transaction processing times. This also results in better resource utilization as it lightens the computational and storage expectations for each node. Along with these benefits, it also offers challenges of maintaining data consistency among shards.  
    \item \textbf{Scalability}: Both blockchain technology and the handling of spatiotemporal data must take into consideration the need to scale efficiently, in order to accommodate an increasing volume of data and volume of users. Scalability becomes all the more critical when these two domains intersect, for instance, blockchain-based spatiotemporal data management, where the volume, velocity, and complexity of the spatiotemporal data presents unique challenges. With the increasing adoption of blockchain, the number of users and transactions grows in an exponential manner. Scalability without can result in congestion, which increases the time taken to transact and the cost of doing so. With the advent of IoT devices, satellites, and mobile applications, spatiotemporal data is produced in huge volume. A single autonomous vehicle, for example, can produce gigabytes of data an hour. With the proliferation of mobile devices and location based services, massive volume of spatiotemporal data is generated. Scalable solutions are needed to process this data in efficient manner. Most of the modern blockchain applications, for example, DeFi, gaming, require near-real-time transaction processing. Scalability ensures that the network will be able to meet these demands without delay.	
    \item \textbf{Energy consumption}
    Blockchain network dealing with spatiotemporal data have to process data that is massive in volume, velocity (real time) and frequency. This makes energy consumption a critical to consider factor. Since spatiotemporal data needs to be updated frequently, therefore intensive computations might be needed by traditional consensus mechanisms. Energy might be consumed more if queries are running on multiple shards.
    \item \textbf{Structure and Index employed}
    Using indices is detrimental for adoption of blockchain in practical real life application, as without its usage query and storage efficiency significantly declines. Various data structures are needed to manage dynamic nature of spatiotemporal data and sequential blockchain transactions. Historical data might need to be referred and processed in a manner that avoids reprocessing of entire chain's data. For example, land registry verification application verifies ownership record (geospatial records) using Merkle tree proofs instead of scanning whole blockchain's data. This reduces storage and processing overhead to a great extent. Further, geo-spatial blockchain enabled ride sharing application needs to locate nearest taxis available using R-tree or Quadtree index without the need to scan whole ledger. Overall, as blockchain's architecture is sequential in nature, an index must be required to run spatiotemporal queries for faster retrieval. Applications that can benefit from this include autonomous vehicle tracking, event tracking, supply chain tracking, crime scene investigation, and distaster management and response system.
    \item \textbf{Security}	
    Blockchain-enabled spatiotemporal environments, which handle data related to both location and time, are essential for various applications such as smart cities, supply chains, autonomous vehicles, and location-based services. Ensuring security is crucial to maintain privacy, trust, and integrity of data. Location must not be tampered to avoid fraudulent activities in many IoT applications. For example, fake location coordinates generated in ride sharing, criminal investigation, and supply chain management related applications. An attacker might forge location based queries and deduce sensitive meeting locations, personal home address, or travel patterns. By utilizing zero-knowledge proofs, decentralized data storage, and homomorphic encryption, a secure spatiotemporal ecosystem can be created by blockchain.
    \item \textbf{Consensus}
    Consensus mechanisms help in guaranteeing integrity of data. Without it, an attacker can claim fake location, spoof a GPS location or alter timestamps in an attempt to initiate fraudulent tracking of delivery. Usually spatiotemporal data needs a validation across regions. For example, shipment being tracked on blockchain travels across various geographical locations. Timestamps must be dealt in consistent manner for different time zones. 
    \item \textbf{Throughput}	
    For effective real time location tracking, high throughput is needed to deal with frequent location updates. Massive streams of continuous data must be stored on blockchain without congestion. High throughput helps in achieving query efficiency and real time analytics.
    \item \textbf{Queries}	
    Queries are important for getting useful insights from spatiotemporal data. Unlike traditional systems, blockchain introduces specific challenges for querying spatiotemporal data due to its decentralization architecture. In real time decision making, queries allow to track vehicles and guides about appropriate traffic routes. With regard to immutable records in blockchain, queries are critical for verification of historical data, specifically in supply chain audits. Queries help to perform analysis like identifying high traffic regions, finding correlation between time and location, and detecting outbreak of some disease in any region. Spatial indexing and querying practices like R-tree index and geohash index help in reducing search space. Table \ref{tab2} outlines the types of queries in spatiotemporal blockchain enabled environment.
\begin{table*}[ht]
    \centering
    \begin{tabular}{|p{2.3cm}|p{12cm}|}
    \hline
       \textbf{Query category}  &  \textbf{Queries in category} \\
       \hline
      Spatial queries  
      & \textbf{Spatial joins}: Find the routes that overlaps with congestion areas.,
      
      \textbf{Point queries}: Find the current location of ambulance., 
      
      \textbf{Nearest neighbor queries}: Find nearest fire stations in emergency situations., 
      
      \textbf{Range queries}: Find the users that resides within 10 km of a shopping mall. \\
        \hline
         Temporal queries  
      & \textbf{Point based queries}: Find the location of ambulance at 8:00 AM on 26th February, 2025., 
      \textbf{Interval based queries}: Find the motion history for a drone between 8:00 AM and 2:00 PM., 
      \textbf{Time series queries}: Track changes in temperature over a month in weather forecasting application. \\
        \hline
         Spatiotemporal queries  
      & \textbf{Event detection queries}: Detect disaster at a specific location., 
      
      \textbf{Join queries}: In a ride sharing application, correlate requests of rides with drivers available within a 10 minute timeframe., 
      
      \textbf{Trajectory queries}: Track the trajectory or route taken by a vehicle in fleet management application. \\
        \hline
         Historical queries  
      & \textbf{Provenance queries}: Find the location history in legal compliance application., 
      
      \textbf{Version queries}: Trace the journey of a product from source location to customer's location. \\
        \hline
    \end{tabular}
    \caption{Type of queries in spatiotemporal blockchain enabled environment}
    \label{tab2}
\end{table*}
    \item \textbf{Latency}
    Latency is the delay occurred when a transaction is submitted till its confirmation on the blockchain. It affects the response time that is critical to maintain in spatiotemporal applications that need instantaneous decisions. For example, emergency response management system, autonomous vehicles. Low latency directly minimizes the consistency issues in location data. The trade-off between throughput and latency must be carefully managed. If throughput is attempted to be increased then this increases the latency too. The consensus mechanisms used in blockchain validation can result in latency. If sharding is implemented then cross sharding communication costs must also be addressed to reduce latency.
    \item \textbf{Consistency}
    Consensus and consistency are related but difference concepts. Consensus is a procedure by which multiple entities in a distributed environment reach a common agreed upon value that is critical for integrity of the system. Consistency is however a property or feature of a system where multiple entities have a same view of the same data item either at the same time (in case of strong consistency) or eventually (for eventual consistent systems). It ensures that data is up to date even when simultaneous updates are made. Consensus in distributed or decentralized environments basically ensures consistency. Spatiotemporal applications have a critical demand for consistency as inconsistent data can result into inaccurate result reporting in disaster management and other safety critical applications. Cross border shipments tracking also needs synchronization in timestamps and location spanning variety of time zones and spatial regions. In smart traffic management systems, inconsistent query results can lead to wrong routing and predictions.
    
    Consensus alone is not sufficient in blockchain enabled spatiotemporal applications. The aspect of quality and accuracy of data is addressed only in consistency as consensus only addresses the agreement concerns. For example smart traffic management application uses consensus mechanisms to ensure that each node in the network agrees on location data of some vehicle. However, it cannot be validated that if the data is inconsistent or outdated. Context dependent nature of spatiotemporal data necessitates to consider contextual correctness of data in terms of accurate location and synchronization of time. That is, timestamp allocated by GPS must be as per the actual event time and data location must reflect location in physical world. Consensus cannot alone verify these things. 
\end{itemize}
\subsection{Feature vector}
For the sake of clarity and understanding, we combined the related features into unified groups and formed a feature vector. The grouping has been defined in table \ref{tab1}.
\begin{table*}[ht]
    \centering
    \begin{tabular}{|p{5cm}|p{5.5cm}|}
    \hline
       \textbf{Group title}  &  \textbf{Features} \\
       \hline
      Architectural Foundations  
      & Structure, Index Employed \\
      \hline
      Performance and Query Optimization & Scalability, Throughput, Latency, Queries \\
       \hline
       Trust and Security Fortifications & Security, Consensus, Consistency\\
        \hline
       Resource efficiency and Optimization & Sharding, Energy Consumption \\
        \hline
    \end{tabular}
    \caption{Feature vectors}
    \label{tab1}
\end{table*}
\section{Comparative analysis: Insights into challenges and benefits}
We assessed the research efforts that have been against the fusion of spatiotemporal data management and blockchain in detail. Comparative analysis of these works have been made with respect to feature vector designed and discussed in the previous section. Tables \ref{tab3}-\ref{tab6} present a comparative view of research contributions for respective feature vectors. Detailed analysis, identification and detailed discussion of challenges along with real life applications, and review of benefits of fusion of spatiotemporal and blockchain data management in the view of feature vectors have been made in the upcoming sections. It is important to note that since features are relevant only for such studies that have presented or proposed some query method, index, or framework relating to blockchain and spatiotemporal data management fusion. We therefore, presented only such studies in comparative view in tables \ref{tab3}-\ref{tab6}.
\begin{table*}[ht]
    \centering
    \begin{tabular}{|p{2cm}|p{8cm}|p{2.5cm}|}
    \hline
    \textbf{Research work}  &   \textbf{Structure}  &  \textbf{Index Employed?} \\
      \hline
  \cite{zheng2024data}  & block structure based on 3D-Merkle tree, DAG &	Yes\\
      \hline
 \cite{hojati2024dstree}   & two level tree structure &	Yes\\
      \hline
 \cite{xing2021blockchain}   & hash pointers to connect subchains &	Yes
\\
      \hline
   \cite{jiang2019cross}    & BigcainDB, IPFS and DAG structure & Yes\\
      \hline
    \cite{chen2020data}   & BigchainDB structure based on hash pointers and non-temper index &	Yes\\
  \hline
    \cite{nasrulin2018robust}   & None	& No\\
      \hline
\cite{qu2019spatio}   & block DAG &	Yes\\
      \hline
\cite{wen2021blockchain}   & Geohash and Unix timestamps &	Yes\\
      \hline
\cite{li2021blockchain}   & None &	No\\
      \hline
\cite{trimborn2018crix}  & None	& Yes\\
      \hline
\cite{liu2019energy}   & None &	No\\
      \hline
\cite{feng2022efficient}   & tree based structures (BB+ tree and I2B+ tree) &	Yes\\
      \hline
\cite{zhang2019gem}   & tree based structure GEM2 tree &	Yes\\
      \hline
\cite{lee2023hierarchical}  & tree based structure HQ-sLSM tree &	Yes\\
      \hline   
    \end{tabular}
    \caption{Comparative analysis with respect to \textbf{Architectural Foundations}}
    \label{tab3}
\end{table*}
\begin{table*}[ht]
    \centering
    \begin{tabular}{|p{1.5cm}|p{1.5cm}|p{2cm}|p{5cm}|p{4cm}|}
    \hline
      \textbf{Research work} & \textbf{Scalability}  &  \textbf{Throughput} & \textbf{Latency} & \textbf{Queries} \\
           \hline
  \cite{zheng2024data}   & Yes & upto 1800 TPS & three-dimensional data queries and multi-block queries & 70 \% lower than  single-chain blockchain solution\\
      \hline
 \cite{hojati2024dstree}    & Yes & Not evaluated & spatiotemporal queries & Not evaluated\\
      \hline
 \cite{xing2021blockchain}    & Yes & Not evaluated & data queries about account subchain & Not evaluated\\
   \cite{jiang2019cross}    & Yes & upto 645 TPS & queries through smart contracts & average latency of 4.77 s\\
      \hline
    \cite{chen2020data}   & Yes & Not evaluated & No specific query types are discussed & Not evaluated\\
      \hline
    \cite{nasrulin2018robust}    & Yes & Not evaluated & Analytical and range queries & Not evaluated\\
      \hline
\cite{qu2019spatio}    & Yes & values for transactions per second (TPS) are not explicitly provided & ball-point queries, bounded k-NN queries, k-NN queries and range queries & values for transactions per second (TPS) are not explicitly provided\\
      \hline
\cite{wen2021blockchain}   & Yes & 100 transactions per second & Close contact tracing queries involving spatiotemporal data & Not evaluated\\
      \hline
\cite{li2021blockchain}    & Yes & Not evaluated & No explicit discussion & Not evaluated\\
      \hline
\cite{trimborn2018crix}  & No explicit discussion & Not evaluated & No explicit discussion & Not evaluated\\
      \hline
\cite{liu2019energy}    & No explicit discussion & Not evaluated & aggregation queries & Not evaluated\\
      \hline
\cite{feng2022efficient}   & No & Not evaluated & single and multi feature retrieval queries & Not evaluated\\
      \hline
\cite{zhang2019gem}   & Yes & Not evaluated & Authenticated range queries & Not evaluated\\
      \hline
\cite{lee2023hierarchical}  & No & Not evaluated & Range and nearest neighbor queries & Not evaluated\\
      \hline   
    \end{tabular}
    \caption{Comparative analysis with respect to \textbf{Performance and Query Optimization}}
    \label{tab4}
\end{table*}
\begin{table*}[ht]
    \centering
    \begin{tabular}{|p{2.5cm}|p{2.5cm}|p{5cm}|p{5cm}|}
    \hline
     \textbf{Research work} &  \textbf{Security}  &  \textbf{Consensus} & \textbf{Consistency} \\
          \hline
  \cite{zheng2024data}  & No explicit discussion & consensus achieved using the Hashgraph algorithm & Yes\\
      \hline
 \cite{hojati2024dstree}   	& Yes & No explicit discussion & No discussion\\
      \hline
 \cite{xing2021blockchain}    	& No & No explicit discussion & No discussion\\
   \cite{jiang2019cross}   	& Yes & Byzantine Fault Tolerance (BFT) consensus algorithm  & Yes\\
      \hline
    \cite{chen2020data}  	& Yes & alliance voting based, Byzantine fault tolerance, proof of work  & Yes\\
      \hline
    \cite{nasrulin2018robust}   	& Yes & Yes & Yes, due to inherent characteristics of permissioned blockchain\\
      \hline
\cite{qu2019spatio}   	& No explicit discussion & No explicit discussion & using cryptographically signed history in block-DAGs\\
      \hline
\cite{wen2021blockchain}   	& Yes & No explicit discussion & No explicit discussion\\
      \hline
\cite{li2021blockchain}   & Yes & data reconstruction-directed consensus mechanism & No explicit discussion\\
      \hline
\cite{trimborn2018crix}  & No & No explicit discussion & No explicit discussion\\
      \hline
\cite{liu2019energy}  	& Yes & No explicit discussion & No explicit discussion\\
      \hline
\cite{feng2022efficient}   	& No & No explicit discussion & No explicit discussion\\
      \hline
\cite{zhang2019gem}   	& No & No explicit discussion & Yes, hash values to ensure integrity\\
      \hline
\cite{lee2023hierarchical}  & No & No explicit discussion & Yes, hash values to ensure integrity\\
      \hline   
    \end{tabular}
    \caption{Comparative analysis with respect to \textbf{Trust and Security Fortifications}}
    \label{tab5}
\end{table*}
\begin{table*}[ht]
    \centering  
    \begin{tabular}{|p{2.5cm}|p{4cm}|p{4cm}|}
    \hline
     \textbf{Research work}  &   \textbf{Sharding}  &  \textbf{Energy consumption} \\
       \hline
  \cite{zheng2024data} &  Blockchain sharding	& No explicit discussion \\
      \hline
 \cite{hojati2024dstree}  &  No explicit discussion	& No explicit discussion \\
      \hline
 \cite{xing2021blockchain}  &  No explicit discussion	& No explicit discussion \\
      \hline
   \cite{jiang2019cross}   &  No explicit discussion	& No explicit discussion \\
      \hline
    \cite{chen2020data}  &  No explicit discussion	& No explicit discussion \\
      \hline
    \cite{nasrulin2018robust}  &  Yes	& No explicit discussion \\
      \hline
\cite{qu2019spatio}  &  No explicit discussion	& No explicit discussion \\
      \hline
\cite{wen2021blockchain}  &  No explicit discussion	& No explicit discussion \\
      \hline
\cite{li2021blockchain}  &  No explicit discussion	& Yes using SCSDA \\
      \hline
\cite{trimborn2018crix}  &  No explicit discussion	& No explicit discussion \\
      \hline
\cite{liu2019energy}  &  Yes & Yes \\
      \hline
\cite{feng2022efficient}  &  No explicit discussion	& No explicit discussion \\
      \hline
\cite{zhang2019gem}  &  No explicit discussion	& No explicit discussion \\
      \hline
\cite{lee2023hierarchical}  &  No explicit discussion	& No explicit discussion \\
      \hline    
    \end{tabular}
    \caption{Comparative analysis with respect to \textbf{Resource efficiency and Optimization}}
    \label{tab6}
\end{table*}
\subsection{Architectural Foundations} 
Comparative view of research works with respect to architectural foundations is presented in table \ref{tab3}. Detailed analysis and comparison in presented in forth coming section; illustrating the challenges faced and benefits achieved. The discussion also discusses the real life applications along.
\subsubsection{Structure and Index Employed}
Considering data structures usage, fusion of blockchain and spatiotemporal data management introduces some challenges and benefits.

Storage overhead: Utilizing index to manage and process frequently changing spatiotemporal data on blockchain can result in exceeding storage capacities as index itself consumes memory.

Complex nature of data: Spatiotemporal data needs specialized structures and indices to handle spatial and temporal aspects effectively. However, blockchains uses specialized data structures that deals with append only nature data. Therefore, these indices cannot be directly applied to handle spatiotemporal data. 

Query difficulties: Spatiotemporal data applications are computationally expensive in nature and will run slow on blockchain as the blockchains cannot handle high performance queries such as "finding a vehicle lying within a 100m range at a specific time". 

Some benefits are also expected such as temper-proof data storage which is critical in applications of environment modeling, supply chain tracking, legal disputes management involving location data. Risk of data manipulation is also reduced due to decentralized  nature of blockchain. 

The paper \cite{zheng2024data} presented a hybrid index that is a mix of tree and block based structure. The proposals \cite{hojati2024dstree}, \cite{feng2022efficient}, \cite{zhang2019gem}, and \cite{lee2023hierarchical} are tree-based. The research works \cite{xing2021blockchain}, \cite{chen2020data}, \cite{wen2021blockchain}
used a hash structure in their design. The proposals \cite{jiang2019cross} and \cite{qu2019spatio} are block-based. Tree-based indices are good at handing range queries and historical lookup queries; useful in geofencing and smart cities applications. However, immutability nature results in high update overhead. Hash based indices results in fast retrievals for exact match queries, effective location based sharding, and low consumption of storage. However, they have a limitation of hash collisions and poor performance for range queries. Such indices are good for quick retrieval when precise location is known e.g. in parcel tracking GPS based application. Block based indices are good for chronological retrieval queries but they suffer from ineffective and slow location tracking in spatial queries. They are useful for temporal queries such as historical analysis of disaster response but slow for real time location tracking. A hybrid approach might be effective for optimized data management. For example, a smart city blockchain can use R-tree for nearest neighbor queries, hash tables for lookups based on vehicle ID, and time based block partitioned index for analyzing traffic congestion and historical trail.
\subsection{Performance and Query Optimization} 
Comparative view of research works with respect to performance and query optimization is presented in table \ref{tab4}. Detailed analysis and comparison in presented in forth coming section; illustrating the challenges faced and benefits achieved. The discussion also discusses the real life applications along.
\subsubsection{Scalability}
Data generation with high frequency: Spatiotemporal applications generate increasingly massive volume of continuous data streams that leave blockchains with declining throughput challenges. For example, fleet management system track millions of vehicles per second which greatly affects the transaction per second capacity of blockchain systems such as Ethereum.

Storage bloating: In contrast to financial transactions that are discrete in nature, spatiotemporal data is continuous in nature and accumulates over time that leads to blockchain storage requirements. For example, a weather forecasting system records environmental aspects such as temperature, wind, and humidity levels at multiple sites each second results in exponential growth in blockchain size.

Latency in real time processing of queries: The primary concern of blockchain networks include security, privacy, and decentralization. Speed is at the second priority that causes delays when processing spatiotemporal data. For example, smart traffic monitoring system may initiate query at blockchain to check for traffic congestion. The query processing may suffers from high latency that makes real time analytics impractical to perform.

Indexing difficulties on immutable blockchain:
Existing blockchain systems don't have a support for traditional spatial indices such as R-trees. That makes spatiotemporal data retrieval inefficient. For example, a ride sharing application finding the vehicles lying inside a 10km range radius might end up querying a larger portion of blockchain. This is in contrast to spatial indexing scenario in which an optimized index helps in reducing search space and reducing complexity and latency of query processing.

Apart from these challenges, there are numerous benefits too. As blockchain supports immutability, location history (spatiotemporal data) can be ensured to be untampered. This is highly valuable for applications where accountability is critical to maintain such as supply chain tracking. The decentralization feature of blockchain helps eliminating single point of failure and risk of data manipulation at some centralize site. This is useful in disaster response management application, where trust in data is critical and important to maintain. Smart contract can be utilized in spatiotemporal data management applications for automation. For instance, in a disaster management application, if a disaster happens in a specific geographical region, insurance claims or payouts might be automatically issued using smart contract. While legacy blockhain networks do not support massive transaction throughput, recently some contributions have been made to address scalability and throughput challenges. Layer-2 scaling and DAG based blockchains such as Hedera are some noticeable solutions in this regard. Layer-2 scaling takes batches of frequent updates and after summarization commits only summarized data on chain. Batches are managed off chain that results in throughput enhancement. DAG aims at replacing block based structure of typical blockchain with graph based structures that supports parallel processing of transactions; thereby increasing throughput. 

All of the proposals discussed in related work section supports scalability, except the works \cite{trimborn2018crix}, \cite{liu2019energy}, \cite{feng2022efficient}, and \cite{lee2023hierarchical}.

\subsubsection{Throughput and Latency}
Spatiotemporal data is generated by sensors with high velocity and frequency. This necessitates the blockchain systems to increase throughput to enhance user experience. Traditional solutions who have adopted blockchain work at low throughput e.g. Ethereum. To apply blockchain concept in spatiotemporal data management, high throughput solutions must be adopted. Latency cannot be bear for spatiotemporal data processing as it has clear real time processing requirements. For example, safety critical applications cannot withstand high latency as it can result in loss of human life. Airline surveillance, autonomous vehicles, and emergency disaster response systems must have a quick response time and low latency.

All the research works except \cite{zheng2024data}, \cite{jiang2019cross}, and \cite{wen2021blockchain} have not evaluated throughput and latency of their solutions. These solutions must be evaluated against throughput and latency to better reflect their suitability to be adopted in real life applications. 
\subsubsection{Queries} 
As blockchain offers a consensus mechanism, delays might be introduced in processing real time queries. Massive spatiotemporal data cannot be stored on chain due to storage limitations. Often, spatiotemporal data comprises of redundant information such as repetitive locations that exhausts the storage capacity. One of the adopted solution is to store the data off chain. But such solutions add complexity to data retrieval process and data validation. Traditional blockchains involve simple queries that usually involves retrieving a specific block. They do not have a native support for spatiotemporal queries that are complex in nature and require non-trivial processing techniques. Data integrity can be challenging as checking the integrity of data off chain needs complex mechanisms such as cryptographic hashes or using Merkle tree based solutions. Further, interoperable framework might be needed to deal with blockchain enabled spatiotemporal data management since spatiotemporal data comes from variety of sources that needs unification. 
\subsection{Trust and Security Fortifications}
Comparative view of research works with respect to trust and security fortifications is presented in table \ref{tab5}. Detailed analysis and comparison in presented in forth coming section; illustrating the challenges faced and benefits achieved. The discussion also discusses the real life applications along.
\subsubsection{Security}
Spatiotemporal data usually takes the form of location histories and trajectories of movement that is sensitive to reveal. Ensuring privacy and security is challenging as applying security techniques will result in performance degradation of the system. Especially achieving validation of spatiotemporal data in real time is challenging. The data usually arrives from varying sources and sensors, possessing variety of formats. Interoperability of these sources is quite challenging. Maintaining security standards helps in retaining control over the data and useful for variety of applications such as supply change management, disaster management or military related applications.

The research works \cite{hojati2024dstree}, \cite{jiang2019cross}, \cite{chen2020data}, \cite{nasrulin2018robust}, \cite{wen2021blockchain}, \cite{li2021blockchain}, and \cite{liu2019energy} have studied security and presented security solution. However, the other studies have not discussed security of the presented solutions. 
\subsubsection{Consensus}
Traditional consensus mechanisms used in blockchain cannot handle spatiotemporal data because if its high volume and dynamic nature. Maintaining consensus on such huge datasets without disturbing performance is a challenge. Further, minimizing the consensus overhead without losing accuracy of data is also difficult as the data often has fine-grained details such as timestamps in milliseconds. 

Proper consensus technique for spatiotemporal data can help in enhancing data integrity and assures that events recorded over time are not tampered. Due to decentralization, consensus techniques helps in transparent sharing of data without involving intermediate parties. Auditing of data also becomes easy as consensus directly enhances accountability that is significant in supply chain tracking applications.

\cite{zheng2024data} presented a DAG based consensus mechanism that supports fair ordering of transactions and high throughput. This protocol is ideal for dynamic nature data such as spatiotemporal data due to its low latency. Also, it is best for applications that have low latency demands and need fair ordering of transactions or events such as traffic management applications or environmental changes tracking application. However, it suffers from communication overhead because of using gossip about gossip mechanism in its working. 

\cite{jiang2019cross} presented BFT based consensus mechanism that supports strong consistency and security in the presence of malicious nodes. It is suitable for consortium blockchains but not admirable for large network as scalability issues may occur when communication among the validators exceed. It can be utilized in spatiotemporal applications that have strong consistency, temporal integrity, and security demands. For example, in military or land registry applications. 

\cite{chen2020data} proposed alliance voting based consensus mechanism in which trusted nodes, known as alliance, vote for transaction validation that results in less consumption of energy and faster finality of transactions. However, collusion might occur among trusted members group. It is best suitable for industry related applications where a group of stakeholders manage the system e.g. smart city infrastructure and have moderate scalability and trust demands. 

Data reconstruction based approach is presented in \cite{li2021blockchain} that relies on reconstructing data from multiple sources. It primary supports data redundancy and high availability. However, this approach is vulnerable to data tampering issues. The data reconstruction process requires massive resources for computation that may introduce latency during validation process. It can be used in applications where data arrive from multiple sources (IoT sensors) and then cross validation happens. For example, environmental monitoring or climate change tracking. Since it supports data integrity, it can be used in disaster recovery applications too.

\subsubsection{Consistency}
In spatiotemporal blockchain enabled environment, achieving consistency in real time is challenging as consensus and propagation of data introduce latency. Straining of blockchain can happen due to enormous amount of data making it difficult to ensure consistency. Providing consistency while preserving data privacy is also difficult. 

Ensuring consistency is not straight forward process. The spatiotemporal data can be outdated or noisy due to sensor error or delay in communication. In such cases, consensus alone cannot validate the data quality. Data cleaning or related techniques might be required in this respect. Consensus mechanisms might introduce latency in the system which is not bearable for safety critical and time sensitive applications. Spatiotemporal data is continuous in nature. Recording each update can bloat the blockchain. 

Blockchain's immutability feature helps in achieving consistency of data (tamper proof movement of goods in supply chain tracking application. Decentralization feature assists in transparent and consistent sharing of data in cross-border logistics management application. Temporal consistency can be effectively enhanced using blockchain's feature of assigning timestamp to transactions and events and ensuring their sequencing. Data provenance can be easily provided by blockchain as it can track changes in data. For example, tracking changes in land registry system. 

All the considered research contributions have discussed consistency or provided consistency solutions. However, the works \cite{hojati2024dstree}, \cite{xing2021blockchain}, \cite{wen2021blockchain}, \cite{li2021blockchain}, \cite{trimborn2018crix}, \cite{liu2019energy}, \cite{feng2022efficient} have not explicitly discussed the consistency and its implications.

\subsection{Resource efficiency and Optimization}
Comparative view of research works with respect to resource efficiency and optimization is presented in table \ref{tab6}. Detailed analysis and comparison in presented in forth coming section; illustrating the challenges faced and benefits achieved. The discussion also discusses the real life applications along.
\subsubsection{Sharding}
Spatiotemporal data has both time and location information that is inherently resource intensive to handle. Sharding when applied to spatiotemporal enabled blockchain environment introduces unique challenges and benefits that are discussed under. 

Cross shard communication: In such environment, historical trail is maintained that requires propagation to ensure that all data is up to date across the shards. Lets assume a blockchain system that tracks delivery vehicles. The data about each vehicle is spatiotemporal in nature (GPS coordinates and timestamps). Upon sharding, each shard handles specific geographical region. Lets assume one shard X is responsible for respective geographical area X. Shard Y is responsible for area Y. Delivery vehicle moves from area X to area Y. In this case stepwise propagation of updates is required across shards. When vehicle begun its journey from area X, data is stored within shard X. As the vehicle moves to area Y, its location now lies under the jurisdiction of shard Y therefore, new location data must be recorded in Y. To maintain consistency across the network, updates in shard Y must be propagated to shard X. This ensures that historical trail is preserved across the shards. 

Cross shard queries: In spatiotemporal systems, queries that deal with data from multiple shards are quite common. For example, queries that retrieve a;; events occurring within 10km radius may need accessing multiple shards that leads to increased latency and complexity if  not managed effectively. Without cross shard propagation of updates, partial or incomplete results might be reported by a query operation. 

Dynamic data distribution: Another challenge is the dynamic nature of spatiotemporal data. Spatiotemporal data is dynamic in nature and needs continuous updates. Sharding must then result into frequent re-sharding and data distribution that significantly increases network and computational overhead.

Security concerns: From security perspective, spatiotemporal data is sensitive in nature as the entity's location data must not be exposed. Sharding exposes this sensitive data to small number of nodes that increases the risk of targeted security attacks and breaches. Ensuring cross shard security and privacy is non-trivial.

Apart from these challenges, sharding also offers some benefits for spatiotemporal blockchain enabled applications such as scalability, efficient query processing, energy efficiency, varying consensus mechanism, fault isolation, and enhanced availability. As sharding distributes the data and load, therefore it is feasible for large applications such as autonomous vehicles, supply chain tracking systems, and smart cities. In such applications it is ideal to divide the load into shards distributed across regions to prevent bottleneck issues. Sharding also assists in query processing. For example, geofencing application can efficiently verifies if a delivery drone is lying inside a restricted space since only relevant region's shard needs to be queried. The solution with sharding incorporated will be energy efficient as the nodes need only to process local transactions instead of processing whole blockchain's transactions. Further, in some scenarios, different shards might need varying consensus mechanisms. For example,  light weight consensus mechanism might be required for the shard dealing with real time traffic monitoring whereas a comparatively more secure consensus mechanism might be required for historical climate data processing shard. Also, if one shard fails, remaining active ones can take over and make the system available. This feature is quite helpful in disaster response system. 

The research works \cite{xiong2020best}, \cite{nasrulin2018robust}, and \cite{zheng2024data} presented and discussed sharding solutions. However, sharding is not evaluated against possible challenges as discussed above. All of the research studies explored apart from these three haven't discussed sharding. Considering the benefits offered by sharding in spatiotemporal environment, it is important to utilize this technology in design of solutions.

\subsubsection{Energy Consumption}
Due to cross shard spatiotemporal queries and frequent update in location, intersection of blockchain and data management introduces some challenges.

Frequent updates: Inherently, spatiotemporal data generated by GPS sensors require frequent update to location in real-time location tracking applications. Using blockchain for dealing with such data increases volume of transactions and induces high energy demands.

Scaling induces high energy consumption: As spatiotemporal data is large in volume and dynamic in nature, it requires scalable solutions for processing and management. Recently, efforts to increase scalability of blockchain have been made such as Layer-2 solutions. But, they introduced the high energy consumption costs. 

Redundant information: More often, there might be low value information associated with spatiotemporal data processing. For example, a repeated location ping by a location based service. Storing such data on blockchain can be very energy wasteful solution. One solution couild be to pre-process the data off chain but it may negatively affects integrity of data.

Complex query processing: Spatiotemporal enabled IoT applications often deals with complex queries such as k-NN and range queries. Unfortunately, such queries are not supported by blockchain system natively. Therefore, extra resources must be required that increasing energy consumption costs.

Apart from these challenges, blockchain offers decentralized management of spatiotemporal data that eliminates dependency on a single server; improving energy efficiency eventually. Token rewards in blockchain environment offers incentives to energy efficient contributors. 

The research work \cite{liu2019energy} and \cite{li2021blockchain} have discussed the energy consumption of solutions and attempted to save energy. Rest of the works haven't discussed this aspect.

\section{Summarized insights}
To conclude, important insights have been summarized and presented as under.

\textbf{Off-chain vs on-chain:}
To make spatiotemporal data management solution scalable, it can be stored off-chain. However, critical data can be handled on-chain.

\textbf{Specialized consensus mechanisms:}
Traditional consensus mechanisms are not suitable for dynamic nature data like spatiotemporal data. Therefore, we need specialized consensus management techniques that suits the dynamic and real time nature of this data.

\textbf{Adaptive partitioning:}
Traditional blockchain assume data shards as independent. However, spatiotemporal data has interdependent components e.g., movement vehicle across shards. Therefore, solutions like temporal sharding and geo-sharding can be adopted to avoid congestion in high traffic areas, improve load balancing and query efficiency. 

\textbf{Energy efficient solutions:}
Energy usage must be monitored and adjusted according to real time data to increase energy efficiency. The concept of geographically load aware consensus can be used in which nodes lying in high traffic regions adopt a low energy consuming protocol such as Proof of Authority, whereas low traffic regions adopt a protocol that consumes adequate energy but ensure security such as Proof of security. Frequent updates can be processed and stored off-chain to save energy consumption. This results in successfully validating spatiotemporal transactions on blockchain while conserving energy.

\textbf{Use of hybrid indexing:}
Traditional blockchains have slow lookups issue as there is no native support of spatiotemporal indexing. Ideally, a multi-indexing approach could be adopted in which multiple structures might be utilized according to the nature of data. For example, spatial lookups can be handled with Rtrees, temporal event based queries can be handled with B-trees, and hash structure can be used for integrity verification.
\section{Conclusion}
Blockchain in combination with spatiotemporal data enhances integrity, security, and traceability of data that is useful in variety of IoT applications including traffic management, supply chain optimization, smart cities. The integration of both the technologies has introduced many challenges that need to be addressed. The existing works in the literature are not detailed and extensive in terms of evaluating the challenges in data management for blockchain enabled spatiotemporal environment. In this paper, we have extensively reviewed the literature in terms of integration of two technological aspects: blockchain and spatiotemporal data management. We designed a feature vector for detailed comparative analysis based on architectural foundations, performance and query optimization, trust and security foundations, resource efficiency and optimization. We assessed the efforts made in the literature related to the fusion of blockchain and IoT (spatiotemporal data in specific) with regard to the designed feature vector. Important insights are presented and challenges are also highlighted. The comparative analysis also incorporated real life applications. By leveraging off-chain and on-chain data management, adaptive partitioning, specialized energy conserving consensus mechanisms, hybrid or multi-indexing, cross shard query optimization, blockchain can be optimized and adopted for spatiotemporal data management.  
\section{Declarations}
All authors declare that they have no conflicts of interest.

\section{Data Availability Statement}
Data sharing is not applicable to this article as no datasets were generated or analyzed during the current study.

\section{Author Contributions Statement}
All authors contributed to the conception and design of the study. Material preparation, data collection, and analysis were performed by the authors. All authors read and approved the final manuscript.

\section{Funding Statement}
This research received no external funding.
\bibliographystyle{elsarticle-num}
\bibliography{ref}

\end{document}